# $Eu_2(SeO_3)_2(SO_4)(H_2O)_2$: Symmetry-Enabled Narrow Optical Linewidths and Microsecond Lifetimes for Quantum Memory Candidates

Uchenna Chinaegbomkpa,[1] Ebube Oyeka,[1] Xudong Huai,[1] Ramesh Kumar,[1] Mingli Liang,[3] Jakoah Brgoch,[3] Hugo Sanabria,[2] and Thao T. Tran*,[1]

[1]Department of Chemistry, Clemson University, Clemson, SC 29634, USA.

[2]Department of Physics and Astronomy, Clemson University, Clemson, SC 29634, USA.

[3]Department of Chemistry, University of Houston, Houston, TX 772045003, USA.

**ABSTRACT:** Stoichiometric materials of Eu(III) offer a promising platform for quantum memories attributable to their unique capability to display a distinctive, nondegenerate $^5D_0 \rightarrow {}^7F_0$ transition, which enables precise mapping of optical quantum states into their energy-level manifold—prerequisite for reliable storage and retrieval on demand. However, placing Eu(III) into chiral polar structures, which are necessary for achieving narrow spectral linewidths and long optical lifetimes, is a daunting task. Here, we discover $Eu_2(SeO_3)_2(SO_4)(H_2O)_2$, a rare Eu(III) material that exhibits chiral polar symmetries encompassing both local and global structures. This unique structure is shaped by an appropriate combination of asymmetric ligands. The chiral structure fosters dipole-dipole interactions and *J*-mixing, as characterized by second-harmonic generation, photoluminescence, and magnetic susceptibility. The broken inversion symmetry is supported by the phase-matching behavior of second-harmonic generation. The $^5D_0 \rightarrow {}^7F_0$ transition is observed at 2.145 eV (578 nm) with a narrow linewidth at 78 K and a microsecond-scale optical lifetime. $Eu_2(SeO_3)_2(SO_4)(H_2O)_2$ also exhibits bright red luminescence with a color purity of 95.63%. The analysis of magnetic susceptibility data using Van Vleck's theory results in an effective magnetic moment of $\rho_{eff}$ = 3.33 $\mu_B$/$Eu^{3+}$ and *J*-mixing. Heat capacity data reveal underlying phonon dynamics in the material. This study demonstrates a pathway to realizing new stoichiometric $Eu^{3+}$ compounds with potential for optically addressable quantum memory applications.

## INTRODUCTION

The uniform distribution of rare-earth emitters in stoichiometric crystals presents significant advantages for achieving high emitter densities and narrow linewidths—both critical parameters for quantum memory technologies.[1-4] Rare earth ions are particularly appealing due to their unique resilience to environmental fluctuations, owing to the lanthanide contraction, which allows the 4f ions to exhibit long optical and spin coherence lifetimes.[5-7] Employing stoichiometric materials in quantum memory devices reduces inhomogeneous broadening by minimizing lattice disorder resulting from local strain, impurities, or defect inclusion, which often occurs in doped systems.[5, 8] In addition, the high concentration of emitters mitigates inefficiencies associated with the repeated storage and retrieval of photonic states during quantum memory operations.[9] Further, the solid-state presentation of crystalline stoichiometric materials also reduces motion-induced dephasing while supporting scalable quantum applications and enhancing compatibility with hybrid system integration.[1] When realized, such coherence properties and high emitter densities enable the achievement of long storage times and high storage efficiencies in quantum memory ensembles.

Although all trivalent rare-earth ions exhibit spin coherence properties advantageous for quantum memory applications,[10-12] $Eu^{3+}$ continues to be of interest due to its distinctive $^5D_0 \rightarrow {}^7F_0$ transition. This transition is forbidden in a free ion but weakly allowed by crystal-field-induced $J$-mixing of $^7F_0$ with $^7F_2$.[13] Experimentally, the $^5D_0 \rightarrow {}^7F_0$ transition is observed as a weak intensity band in low-symmetry, non-cubic crystallographic point groups, such as those without an inversion center, including $C_n$, $C_{nv}$ and $C_s$.[14, 15] In addition, both naturally occurring isotopes of europium, $^{151}Eu$ and $^{153}Eu$, possess a nuclear spin of I = 5/2, resulting in a hyperfine structure with energy splittings ranging between 10 and 100 MHz.[3] Notably, the $^5D_0 \rightarrow {}^7F_0$ transition is unique among the $f$-$f$ transitions of the rare earths as both the ground state and the excited states are non-degenerate. Consequently, at cryogenic temperatures and in systems characterized by narrow inhomogeneous linewidths, the hyperfine states can be spectrally resolved. The utility of the $^5D_0 \rightarrow {}^7F_0$ transitions in quantum memories derives from the ability to map optical quantum states into their hyperfine manifold, enabling reliable storage and retrieval on demand.[16] To be effective, the encoded quantum state must be stored for longer times than the optical transmission of the measurement channel. Such long-lived hyperfine states have been demonstrated in $EuCl_3 \cdot 6H_2O$, which displays an ultranarrow inhomogeneous linewidth of 25 MHz at 4 K in isotopically enriched samples for the $Cl^-$ ion, and a coherence lifetime of 0.74 ms upon deuterium enrichment.[8, 17, 18] The utility of $EuCl_3 \cdot 6H_2O$ is, however, constrained by its instability under both air and vacuum conditions, necessitating further development of alternative candidate materials. Recent studies have demonstrated helpful design protocols and enthalpy consideration for predicting stable and stoichiometric $Eu^{3+}$ quantum memory candidates.[19] $Eu^{3+}$ metal-organic frameworks, polymorphs, molecular crystals and compounds consisting of mononuclidic elements serve as promising platforms for quantum memory.[5, 8, 20] Here, we report a new chiral polar and air-stable material, $Eu_2(SeO_3)_2(SO_4)(H_2O)_2$, created using our strategy of placing $Ln^{3+}$ ions in a noncentrosymmetric lattice supported by three asymmetric building units, $(SeO_3)^{2-}$, $(SO_4)^{2-}$, and $H_2O$.[21] This method leverages the structural and electronic properties of the ligands to create a stoichiometric phase with a single polar cationic site occupancy, a relatively large Eu-Eu separation (>3.95 Å) and a narrow $^5D_0 \rightarrow {}^7F_0$ transition, along with high optical density. Using spectroscopic analysis, photoluminescence studies, phase-matching second-harmonic generation (SHG), and magnetization measurements as sensitive probes, we characterize the local environment of the $Eu^{3+}$ ions, the peak energy, linewidth, and lifetimes of the optical transitions. We also report on the thermodynamic properties of the material using heat capacity measurements and employ DFT calculations to understand the nature of orbital overlaps that govern physical behavior. For quantum memory applications, isotopic purification enhances the realization of narrow inhomogeneous linewidths. Here, we report $Eu_2(SeO_3)_2(SO_4)(H_2O)_2$ in the natural abundance of the constituent elements. But, while selenium may have significant variable abundance, $^{16}O$ accounts for 99.76% of oxygen's natural abundance and $^{32}S$ is 94.93% of natural Sulfur.[22] It is also noteworthy that all three chalcogens are nuclear spin-free.

## EXPERIMENTAL SECTION

**Reagents.** Thiophene-2-carbonyl chloride, aniline, KSeCN, $Eu(NO_3)_3 \cdot 6H_2O$, $HNO_3$ (67%), acetone.

**Synthesis of $C_{12}H_{10}N_2OSSe$ (TAS).** The synthesis and characterization of TAS have previously been reported by us.[23]

**Synthesis of $Eu_2(SeO_3)_2(SO_4)(H_2O)_2$.** The synthesis of $Eu_2(SeO_3)_2(SO_4)(H_2O)_2$ was achieved using a similar reaction procedure as those of the already reported Sm, Dy, and Yb analogs, with a modification of the solvent molarity and reaction time.[21] $Eu(NO_3)_3 \cdot 6H_2O$ (2 mmol, 0.892 g) was dissolved in 3 M $HNO_3$ (10 mL) in a 23 mL polytetrafluoroethane(PTFE)-lined autoclave and TAS (2 mmol, 0.62 g) was added. The autoclave was heated to 200 °C at a rate of 30 °C/hour, allowed to dwell for 30 hours, and then cooled slowly to 25 °C at a rate of 3 °C/hour. $Eu_2(SeO_3)_2(SO_4)(H_2O)_2$ was isolated as colorless crystals, washed with deionized water, and dried in air (Yield (0.336 g) = 99% based on $Eu^{3+}$).

**Single-crystal X-ray diffraction.** Single crystal crystallographic data of $Eu_2(SeO_3)_2(SO_4)(H_2O)_2$ were collected at $T$ = 296(2) K on a Brüker D8 Quest diffractometer equipped with a Photon III detector using Mo $K_\alpha$ radiation ($\lambda$ = 0.71073 Å). Data processing (SAINT), scaling (SADABS), structure solution by intrinsic phasing (SHELXT) and refinement by full-matrix least-squares techniques on $F^2$ (SHELXL) were performed using the Apex4 software suite.[24] Hydrogen atoms attached to water molecules were identified from the difference electron density map and refined with appropriate distance restraints. All atoms except hydrogen were refined anisotropically. The crystal structure was viewed with VESTA.[25] The structure solution model and space group were verified using Platon software.

**Powder X-ray Diffraction.** Powder X-ray diffraction data of $Eu_2(SeO_3)_2(SO_4)(H_2O)_2$ were collected using a Rigaku Ultima IV in the 2$\theta$ range of 5 ° – 80 ° at a 0.1 °/min scan rate. Rietveld refinement of the diffraction pattern was performed using TOPAS Academic V6.[26]

**Thermal Analysis.** Thermogravimetric (TG) analysis and differential scanning calorimetry (DSC) measurements were performed using a TA SDT Q600 Instrument using ~10 mg of $Eu_2(SeO_3)_2(SO_4)(H_2O)_2$ powder (heating rate = 20 °C/min, $T$ = 30 °C – 1000 °C, gas = $N_2$, flow rate: 100 mL/min).

**Infrared Spectroscopy.** Infrared spectra of $Eu_2(SeO_3)_2(SO_4)(H_2O)_2$ were collected using the Attenuated total reflection Fourier transform infrared (ATR-FTIR) mode of a Shimadzu IR Affinity-1S in the 500 to 4000 cm$^{-1}$ frequency range.

**UV-Vis-NIR Spectroscopy.** UV-Vis-NIR spectra of $Eu_2(SeO_3)_2(SO_4)(H_2O)_2$ were collected using an Agilent Cary UV-Vis (NIR) 7000 Spectrophotometer via the transmission mode. A pelletized sample was prepared using 48 mg KBr and 2 mg $Eu_2(SeO_3)_2(SO_4)(H_2O)_2$. KBr was dried at $T$ = 200 °C for 24 hours before use.

**Photoluminescence.** A thin layer of $Eu_2(SeO_3)_2(SO_4)(H_2O)_2$ was prepared on a quartz substrate (10 x 10 x 0.1 mm) by the drop-casting method using a methanol suspension of the sample. The suspension was prepared using 7 mg of the sample. The temperature-dependent steady-state measurements at 78 K ≤ $T$ ≤ 400 K were performed using a Horiba Nanolog spectrofluorometer system (HORIBA FL-1039) equipped with a Xenon lamp and a CCD detector (HORIBA Symphony II). The instrument was operated at a 0.1 s integration time, a 1 nm monochromator slit width and thermally controlled using a Lake Shore temperature controller (Model 325). Data collection was achieved using a 1.0 optical density filter fitted along the emitted beam path. The sample was mounted on a cryostat evacuated to $10^{-3}$ Torr while cooling was achieved using liquid $N_2$. The time-resolved photoluminescence measurement was achieved using a home-built setup comprising a PicoQuant TimeHarp 260 TCSPC controller, a Horiba NanoLED pulsed light source (375 nm ± 10 nm) and a Horiba C2 NanoLED driver.

**Second Harmonic Generation (SHG).** Powder SHG measurements were performed on a modified Kurtz nonlinear optical system using a pulsed Nd:YAG laser with a wavelength of 1064 nm. A detailed description of the equipment and methodology has been published.[27, 28] SHG response as a function of particle size was measured on $Eu_2(SeO_3)_2(SO_4)(H_2O)_2$ powder with the following particle sizes: 15±5, 32.5±12.5, 54±9, 69±6, 82.5±7.5, 107.5±17.5, and 137.5±12.5 µm. Relevant comparisons were made with $KH_2PO_4$, a benchmark SHG material at 1064 nm. No index-matching fluid was used in any of the experiments.

**Magnetization and Specific Heat.** DC magnetization measurements on $Eu_2(SeO_3)_2(SO_4)(H_2O)_2$ polycrystalline sample were performed with the Vibrating Sample Magnetometer (VSM) option of Quantum Design Physical Properties

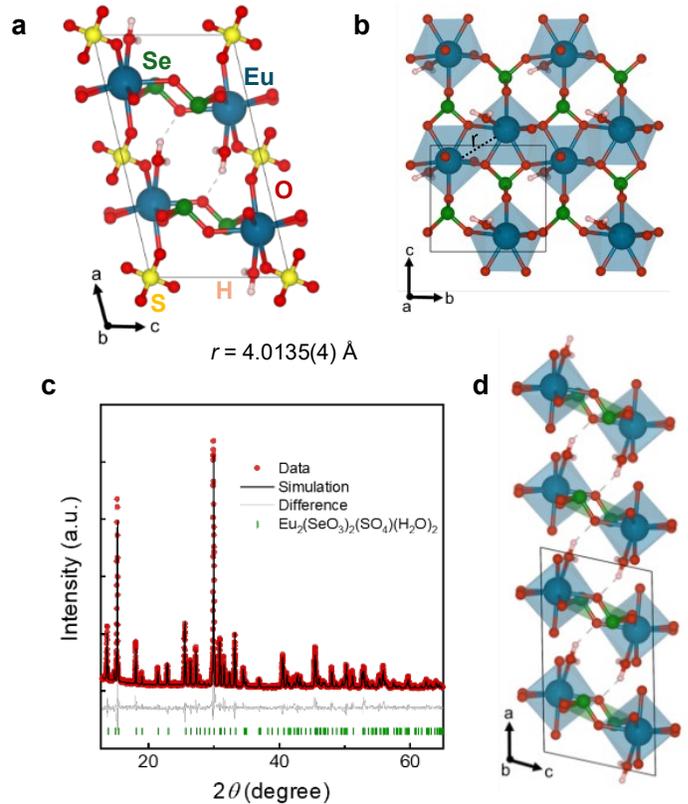

**Figure 1. (a)** Crystal structure of $Eu_2(SeO_3)_2(SO_4)(H_2O)_2$ showing the $EuO_7(H_2O)$, $(SeO_3)^{2-}$, and $(SO_4)^{2-}$ units. **(b)** Minimum Eu-Eu separation along the $b$-axis. **(c)** Rietveld refinement of powder XRD data for $Eu_2(SeO_3)_2(SO_4)(H_2O)_2$. **(d)** Alignment of the $EuO_7(H_2O)$ polyhedra along the $b$-axis, giving rise to the macroscopic electric polarization.

Measurement System (PPMS). Heat capacity was also measured using the PPMS using the semi-adiabatic pulse technique at 2 K ≤ $T$ ≤ 300 K.

**DFT Calculations.** The density of states calculations for $Eu_2(SeO_3)_2(SO_4)(H_2O)_2$ were performed using the Quantum Espresso code with pseudopotentials from the Quantum Espresso pseudopotentials database.[29] The exchange and correlation energies were treated within density functional theory supplemented with a Coulombic interaction (DFT+U), using the Perdew-Burke-Ernzerhof (PBE) generalized gradient approximation.[30] The self-consistency calculations were carried out using a 4 × 4 × 4 k mesh in the irreducible Brillouin zone. The onsite Coulomb repulsion parameter, $U$ for Eu was set as 5 eV.[31] Further chemical bond analysis for crystal orbital Hamilton population (COHP) and crystal orbital bonding index (COBI) were performed using the LOBSTER code.[32]

## RESULTS AND DISCUSSION

**Structure and material design.** $Eu_2(SeO_3)_2(SO_4)(H_2O)_2$, isostructural with the already reported $Ln^{3+}$ analogs (Ln = Sm, Dy, Yb), displays a noncentrosymmetric polar chiral structure in the $C$2 space group, which is rare (**Figure 1, Table S1 and S2**).[21] The structure of $Eu_2(SeO_3)_2(SO_4)(H_2O)_2$ can be described as a 3-D network of distorted $EuO_7(H_2O)$ bicapped trigonal prisms connected through $(SeO_3)^{2-}$ trigonal pyramids and $(SO_4)^{2-}$ tetrahedra units. Each $Eu^{3+}$ ion is bonded to five oxygen atoms from the $(SeO_3)^{2-}$ group, two oxygen atoms from the $(SO_4)^{2-}$group, and one from the $H_2O$ molecule, rendering the local chiral $C_1$ structure of $Eu^{3+}$. The Eu-O bond lengths range from 2.375(15) Å to 2.456(14) Å **(Table S3)**. The $Eu^{3+}$ sublattice features a 1-D zig-zag chain along the $b$-axis with the nearest Eu-Eu distance, $r$ = 4.0135(4) Å **(Figure 1b)**. Placing $Eu^{3+}$ in the chiral $C_1$ local structure with such a significant Eu–Eu separation is a prerequisite for achieving narrow optical linewidths and long spin coherence times.

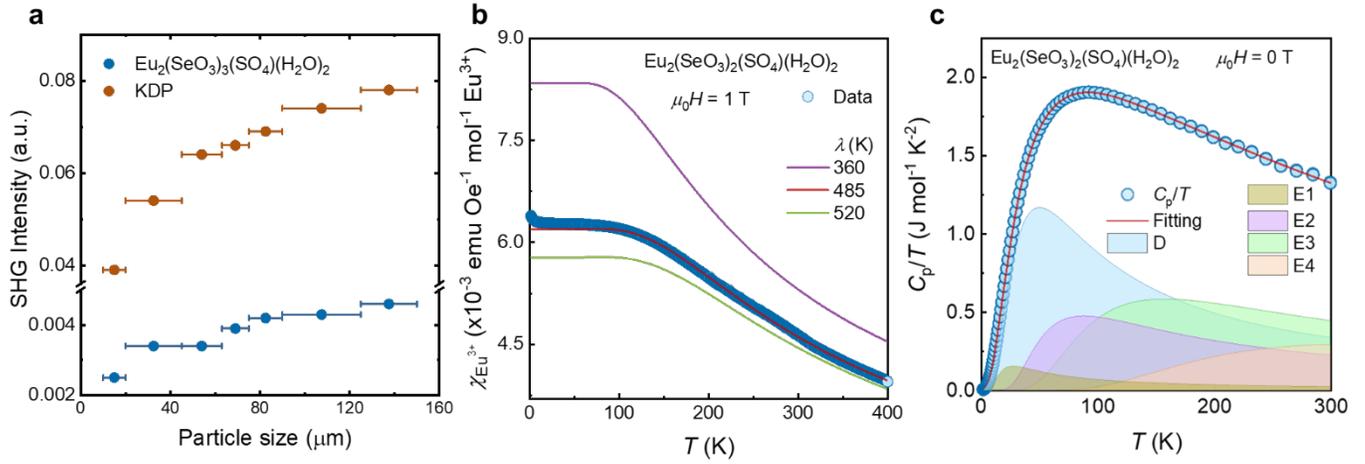

**Figure 2. (a)** SHG signal of $Eu_2(SeO_3)_2(SO_4)(H_2O)_2$ and commercial standard KDP as a function of particle size. **(b)** Temperature dependence of $\chi$ for $Eu_2(SeO_3)_2(SO_4)(H_2O)_2$ at 2 K ≤ $T$ ≤ 400 K and $\mu_0H$ = 1 T The solid blue circles represent the experimental data, and the solid curves show the simulated $\chi$ for different values of $\lambda$. The best fit is given by $\lambda$ = 485 K for 100 K ≤ $T$ ≤ 400 K. **(c)** Molar heat capacity ($C_p/T$) vs temperature ($T$) curve and phonon analysis of $Eu_2(SeO_3)_2(SO_4)(H_2O)$. Phonons were best modelled using a combination of one Debye and four Einstein modes (equation 6).

The chirality of Eu is extended from the $C_1$ point group of Eu to the global structure in the $C2$ space group. The preservation of chiral structures is made possible by a combination of three asymmetric building units: $(SeO_3)^{2-}$, $(SO_4)^{2-}$, and $H_2O$. The Se-O bond lengths range from 1.647(9) Å to 1.698(14) Å, and the O-Se-O bond angles are between 92.9(7)°–104.0(9)°, reducing the symmetry of the $(SeO_3)^{2-}$ group from $C_{3v}$ to $C_1$. The $(SO_4)^{2-}$ group comprises S-O bond lengths ranging from 1.465(10) Å to 1.487(14) Å. The $(SO_4)^{2-}$ tetrahedron is distorted to a pseudo-$T_d$ symmetry with bond angles ranging from 110.0(7) ° to 110.5(7) °. The $H_2O$ molecule bonds to the $Eu^{3+}$ center through its O atom, while forming Se–O–H hydrogen bonds with one O atom of $(SeO_3)^{2-}$. The hydrogen bond reduces the symmetry of $(SeO_3)^{2-}$, as demonstrated in its aforementioned bonding analysis. The H-O bond length and H-O-H bond angle were fixed to 0.91(14) Å and 106(17) °, respectively, during the least-squares refinement, retaining the $C_{2v}$ local symmetry of the $H_2O$ ligand. $Eu_2(SeO_3)_2(SO_4)(H_2O)_2$ starts losing the water molecules at ~300 °C and then decomposes at 700 °C **(Figure S1)**. The local symmetries of the $(SeO_3)^{2-}$, $(SO_4)^{2-}$, and $H_2O$ building units are further substantiated by infrared spectroscopic measurements.

**Infrared spectroscopy.** The ATR-FTIR spectrum of $Eu_2(SeO_3)_2(SO_4)(H_2O)_2$ reveals the local symmetries of the building units. A $(SeO_3)^{2-}$ group adopting the $C_{3v}$ point group would typically display four vibrational modes, comprising two non-degenerate modes, $v_1$ and $v_2$, and two doubly degenerate modes, $v_3$ and $v_4$. Lowering the symmetry breaks the degeneracy of the doubly degenerate modes, resulting in six bands in the vibrational spectrum.[33] In $Eu_2(SeO_3)_2(SO_4)(H_2O)_2$, distortions in bond lengths and angles of the $(SeO_3)^{2-}$ group lower the symmetry to $C_1$. This is evident in the coupled vibrational modes of $v_1(SeO_3)^{2-}$ and $v_3(SeO_3)^{2-}$ observed between 590 – 840 cm$^{-1}$ in the infrared spectrum. For a $(SO_4)^{2-}$ ion, the intramolecular normal modes, $v_1$, $v_2$, $v_3$, $v_4$, corresponding to the $T_d$ irreducible representation, $\Gamma_{Td}$ = $A_1$ + $E$ +2$T_2$ are observed.[34] In $Eu_2(SeO_3)_2(SO_4)(H_2O)_2$, the pseudo-$T_d$ symmetry of the $(SO_4)^{2-}$ group accounts for the vibrational bands at 981 – 1277 cm$^{-1}$, corresponding to the $v_1(SO_4)^{2-}$ and $v_3(SO_4)^{2-}$ modes.[34] The $H_2O$ ligand exhibits three absorption bands in IR spectra, consistent with the $C_{2V}$ point group, $\Gamma_{vib.}$ = 2$A_1$ + $B_1$. The broad bands at 3212 - 3475 cm$^{-1}$ correspond to the characteristic O-H stretching modes of water, while the single band at 1636 cm$^{-1}$ indicates the H–O–H bending mode. The significant increase (blue shift) of the vibrational frequency of the O–H–O bending mode from the characteristic energy, ~1590 cm$^{-1}$ to 1636 cm$^{-1}$, and its relatively greater absorbance compared to the O–H stretching frequencies can be ascribed to the influence of the strong hydrogen bonding in $Eu_2(SeO_3)_2(SO_4)(H_2O)_2$.[35]

**Second Harmonic Generation.** To probe the broken inversion symmetry of $Eu_2(SeO_3)_2(SO_4)(H_2O)_2$ and evaluate its frequency-doubling capability, we performed powder second harmonic generation (SHG) measurements as a function of particle size using 1064 nm radiation **(Figure 2a)**. The SHG of $Eu_2(SeO_3)_2(SO_4)(H_2O)_2$ reveals its ability to convert 1064 nm to 532 nm, doubling the frequency (halving the wavelength). The signal intensity is about 0.06 × $KH_2PO_4$ (KDP) in the 32.5±12.5 μm particle range, comparable to those of the $Ln_2(SeO_3)_2(SO_4)(H_2O)_2$ compounds (Ln = Sm, Dy, Yb).[21] The low SHG intensity of the material can be attributed to the arrangement of the anion groups and the narrow bandwidth between the $f$ states of $Ln^{3+}$ and the $s$ and $p$ states of the ligands **(Figure S3)**. The SHG intensity of the material increases with increasing particle size and reaches a plateau **(Figure 2a)**. This behavior indicates that the material possesses type-1 phase-matching behavior and belongs to the class A category of SHG materials.[28] The phase-matching property of the Eu compound reveals that the fundamental and harmonic waves travel at the same propagation speed—a prerequisite for effectively generating coherent light through the SHG process.

**Magnetization.** We performed temperature-dependent magnetization measurements at $\mu_0H$ = 0.1 T and 2 K ≤ $T$ ≤ 300 K to gain further insights into the spin–orbit ($LS$) coupling, $J$-mixing and crystal field splitting observed in photoluminescence. The magnetic susceptibility was approximated as $\chi \approx M/\mu_0H$. In principle, the $^7F_0$ ground state of $Eu^{3+}$ is non-magnetic ($S$ = 3, $L$ = 3, $J$ = 0). But from the $\chi(T)$ curve of $Eu_2(SeO_3)_2(SO_4)(H_2O)_2$, temperature-independent Van Vleck paramagnetism is observed. From Van Vleck's theory of electric and magnetic susceptibility, a consideration of the population of the higher $^7F_J$ ($J$ = 1 – 6) levels other than the non-degenerate $^7F_0$ ground state is very crucial in the paramagnetic susceptibility analysis of $Eu^{3+}$.[36-38] The spin–orbit interaction $\lambda \mathbf{L} \cdot \mathbf{S}$ relates to the energy levels in $LS$ coupling, such that the energy difference between two states is

$$E_J - E_{J-1} = \lambda J \quad (1)$$

where the magnitude of $\lambda$ indicates the energy difference between the non-magnetic ground state, $^7F_0$, and the first excited state, $^7F_1$.

The temperature-dependent magnetic susceptibility of $Eu_2(SeO_3)_2(SO_4)(H_2O)_2$ was analyzed based on evidence that the $Eu^{3+}$ ion is subject to crystal field effects, as evidenced by the Stark energy levels observed in photoluminescence emission spectra. Using Van Vleck's theory,[38] the explicit paramagnetic susceptibility of the $Eu^{3+}$ ions can be described as

$$\chi_{para}(free\ Eu^{3+}) = \frac{N\mu_B^2}{Z}\left(\frac{A}{3\lambda}\right) \quad (2)$$

where:

$$Z = 1 + 3e^{-\lambda/K_BT} + 5e^{-3\lambda/K_BT} + 7e^{-6\lambda/K_BT} + 9e^{-10\lambda/K_BT} + 11e^{-15\lambda/K_BT} + 13e^{-21\lambda/K_BT} \quad (3)$$

$$A = 24 + \left(13.5\frac{\lambda}{K_BT} - 1.5\right)e^{-\lambda/K_BT}$$
$$+ \left(67.5\frac{\lambda}{K_BT} - 2.5\right)e^{-3\lambda/K_BT}$$
$$+ \left(189\frac{\lambda}{K_BT} - 3.5\right)e^{-6\lambda/K_BT}$$
$$+ \left(405\frac{\lambda}{K_BT} - 4.5\right)e^{-10\lambda/K_BT}$$
$$+ \left(742.5\frac{\lambda}{K_BT} - 5.5\right)e^{-15\lambda/K_BT}$$
$$+ \left(1228.5\frac{\lambda}{K_BT} - 6.5\right)e^{-21\lambda/K_BT} \quad (4)$$

From the foregoing, only $\lambda$ is unknown. Hence, its value can be extracted by calculating susceptibility with different values of $\lambda$ and comparing the calculated results with the experimental temperature-dependent magnetic susceptibility. The magnitude of $\lambda$ inversely relates to the extent of hybridization of the excited states with the non-magnetic ground state.[39] **Figures 2b and S4** present our simulated susceptibility by varying the value of $\lambda$ in equation (2). Our best fit is obtained with $\lambda$ = 485 K (41.8 meV). This demonstrates that $Eu^{3+}$ in $Eu_2(SeO_3)_2(SO_4)(H_2O)_2$ is significantly influenced by spin-orbit coupling, allowing for $J$-mixing. The upturn at $T$ < 5 K could originate from $Eu^{2+}$ impurities.[40, 41] The $\lambda$ = 485 K value of $Eu_2(SeO_3)_2(SO_4)(H_2O)_2$ is comparable to those reported for other relevant $Eu^{3+}$ systems. The $\lambda$ values for $EuBO_3$, $EuF_3$, $Eu_2O_3$, and $EuVO_4$ were reported to be 471 K, 490 K, 460 K, and 560 K, respectively.[39,40] The larger $\lambda$ indicates greater energy splitting of $^7F_0 \to {}^7F_1$ and a lesser degree of intermixing of the $^7F_J$ ($J$ = 0 – 6) manifold. Additionally, the value of $\lambda$ determines the magnitude of the temperature-independent magnetic susceptibility, which is observed as a plateau-like region in the $Eu^{3+}$ susceptibility curve and the corresponding temperature range. The larger the $\lambda$ value, the higher the onset temperature of the temperature-independent paramagnetism and the smaller its magnitude.

We estimate the induced magnetic moment using the relation in (equation 6):

$$\rho_{eff} = \left(\frac{3\chi K_BT}{N\mu_B^2}\right)^{1/2} \quad (5)$$

Where:

$\chi$ = 4.62 (x$10^{-3}$ emu $Oe^{-1}$ $mol^{-1}$) at $T$ = 300 K.

$N$ is the number of $Eu^{3+}$ ions, $k_B$ is the Boltzmann constant in erg $K^{-1}$. $\mu_B$ is the Bohr magneton in erg $gauss^{-1}$. The moment, $\rho_{eff}$ for the $Eu^{3+}$ ion is then calculated to be 3.33 $\mu_B$, which is comparable to other reported $Eu^{3+}$ compounds.[39] To evaluate how magnetization evolves as a function of applied magnetic fields, we measured field-dependent magnetization $M(H)$ **(Figure S5)**. The $M(H)$ curve of $Eu_2(SeO_3)_2(SO_4)(H_2O)_2$ at $T$ = 2 K shows a linear behavior and no saturation at $\mu_0H$ = 0 – 9 T on a linear scale. On a log scale, we observe a very short-range, non-linear behavior that saturates at 0.02 T **(inset in Figure S5)**. Owing to the $J$-mixing and the non-magnetic ground state with $J$ = 0, the Brillouin function calculation for the paramagnetic $Eu^{3+}$ spins is not applicable.

**Heat Capacity.** To study the thermodynamic behavior of $Eu_2(SeO_3)_2(SO_4)(H_2O)_2$, we performed specific heat measurements at $T$ = 2 K – 300 K and $\mu_0H$ = 0 T, 3 T, and 6 T**.** There were no observable differences in the $C_P/T$ versus $T$ plot at the different fields. **(Figure S6)**. In contrast to the Ln = Sm, Dy, and Yb siblings, which showed an anomaly at $T$ = 92 K ascribed to structural order-disorder of the $H_2O$ ligand, $Eu_2(SeO_3)_2(SO_4)(H_2O)_2$ shows no structural phase transition. This is likely due to the stabilization of the $EuO_7(H_2O)$ polyhedra by the inter-layer Eu-O-H---$(SeO_3)^{2-}$ hydrogen bond, which was not observed in the other analog materials.[21] Since there are no structural or magnetic phase transitions in $Eu_2(SeO_3)_2(SO_4)(H_2O)_2$, the total heat capacity is attributed to phonon contribution. The phonon estimation was done by combining one Debye and four Einstein modes in the $Cp/T$ versus $T$ plot **(Figure 2c)**. Equation 6 shows the fit model, while equations 7 and 8 represent the Debye and Einstein models, respectively.[42]

$$C_P(T) = C_D(\theta_D, s_D, T) + C_{E1}(\theta_{E1}, s_{E1}, T) + C_{E2}(\theta_{E2}, s_{E2}, T) + C_{E3}(\theta_{E3}, s_{E3}, T) + C_{E4}(\theta_{E4}, s_{E4}, T) \quad (6)$$

$$C_D(\theta_D, T) = 9 s_D R \left(\frac{T}{\theta_D}\right)^3 \int_0^{\theta_D/T} \frac{(\theta_D/T)^4 e^{\theta_D/T}}{[e^{(\theta_D/T)}-1]^2} \, d\frac{\theta_D}{T} \quad (7)$$

$$C_E(\theta_E, T) = 3 s_E R \left(\frac{\theta_E}{T}\right)^2 \frac{e^{\theta_E/T}}{[e^{(\theta_E/T)}-1]^2} \quad (8)$$

be attributed to the hydrogen atoms, whose vibrations may contribute the least to the oscillator count. The four Debye oscillators describe the four lattice subunits corresponding to the Eu sublattice and the $(SeO_3)^{2-}$, $(SO_4)^{2-}$ and $H_2O$ functional groups. The Einstein oscillators ($s_{En}$ and $\theta_{En}$, n = 1 - 4) could describe the three functional groups, $(SeO_3)^{2-}$, $(SO_4)^{2-}$ and $H_2O$ where the first and second Einstein modes could be attributed to frequencies arising from the water molecules, while the third and fourth Einstein modes describe the $(SO_4)^{2-}$ and $(SeO_3)^{2-}$ groups, respectively. At our experimental conditions, the specific heat of

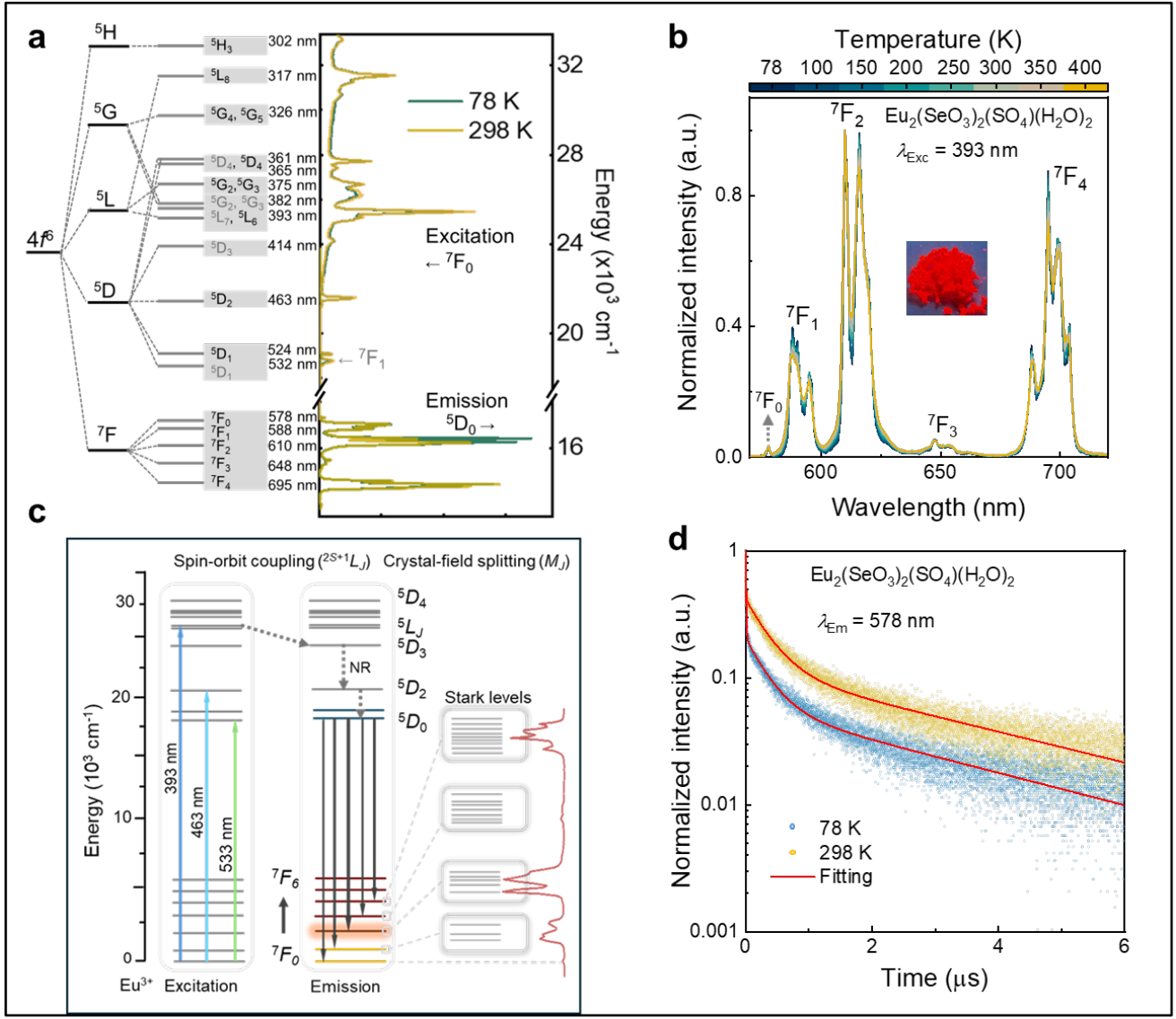

**Figure 3.** (a) Excitation and emission spectra of $Eu_2(SeO_3)_2(SO_4)(H_2O)_2$ at $T$ = 78 K and 296 K. (b, c) Temperature-dependent emission spectra at 78 K ≤ $T$ ≤ 400 K and proposed relaxation mechanism and energy level diagram of $Eu_2(SeO_3)_2(SO_4)(H_2O)_2$. The inset in (3b) is the photoluminescence emission of $Eu_2(SeO_3)_2(SO_4)(H_2O)_2$ under UV (254 nm) irradiation. (d) Photoluminescence lifetime of the $^5D_0 \rightarrow {}^7F_0$ transition of $Eu_2(SeO_3)_2(SO_4)(H_2O)_2$ at 78 K and 298 K using 375 nm excitation and emission at 578 nm.

where $\theta_D$ is the Debye temperature; $\theta_{E1}$, $\theta_{E2}$, $\theta_{E3}$ and $\theta_{E4}$ are the Einstein temperatures; $s_D$, $s_{E1}$, $s_{E2}$, $s_{E3}$ and $s_{E4}$ are the oscillator strengths; and $R$ is the molar gas constant. The resulting parameters are presented in **Table S4**. The sum of the oscillators, 20.22(47), is closely comparable to the expected value of 21, which is the total number of atoms per formula unit of $Eu_2(SeO_3)_2(SO_4)(H_2O)_2$. The difference could

$Eu_2(SeO_3)_2(SO_4)(H_2O)_2$ does not show any saturation up to 300 K, captured by the fourth Einstein mode whose population continues to increase up to 300 K. $Eu_2(SeO_3)_2(SO_4)(H_2O)_2$ also does not show the Schottky anomaly observed in $Ln_2(SeO_3)_2(SO_4)(H_2O)_2$ (Ln = Sm, Dy, Yb) due to its non-magnetic ground state with $J$ = 0 (**Figure S6**).

**UV-Vis-NIR spectroscopy.** The absorption spectrum **(Figure S8)** was collected to investigate the interaction of the orbital $L$ and spin angular momentum $S$ ($LS$ coupling) and level of the ground state, $^7F_0$, and its higher $J$ levels, $^7F_J$ ($J = 1 – 6$), are clearly presented in **(Figure S8a)** while the whole spectrum (350 – 2500 nm) is presented as a support for the photoluminescence properties in **(Figure S8b)**. Band assignments are consistent with literature values for $Eu^{3+}$.[43] The low-intensity peaks lying between the $^7F_0 \rightarrow ^7F_J$ ($J = 1 - 3$) could be due to splitting of the Stark levels. The absorption spectrum also indicates strong spin-orbit ($L+S$) coupling of the $Eu^{3+}$ cations and coupling with the phonon modes of the anions. The coupling is evident from the sidebands surrounding the $^7F_0 \rightarrow ^5D_1$ transition and the band broadening of the $^7F_0 \rightarrow ^5D_3$ transition. The most intense peak is the $^7F_0 \rightarrow ^5L_6$ transition at 395 nm (3.139 eV). Only the zero-phonon line of the $^7F_0 \rightarrow ^5D_2$ is observed, while no peaks were observed for the $^7F_0 \rightarrow ^5D_0$ transition. The broad band, however, at 588 nm is attributed to the $^7F_1 \rightarrow ^5D_0$ transition from the thermally populated $^7F_1$.

**Photoluminescence.** Compared to other non-Kramers ions, $Eu^{3+}$ holds a unique advantage in that the dominant emitting state of the $^5D_J$ levels and the lowest level of the $^7F_J$ manifold are non-degenerate ($J = 0$).[44] In this study, we used photoluminescence to study the $Eu^{3+}$ site symmetry and the optical linewidths and lifetimes of the spectral transitions. Having a $C2$ space group and a $C_1$ point symmetry for the $Eu^{3+}$ ion, all $^5D_0 \rightarrow ^7F_J$ ($J = 0 – 4$) transitions are expected to be observed. Using an excitation energy of 3.139 eV ($\lambda = 393$ nm), the symmetry-forbidden $^5D_0 \rightarrow ^7F_0$ transition is observed at 2.145 eV (578 nm) as a single narrow band, supporting the design strategy for chiral polar symmetry of $Eu^{3+}$ in $Eu_2(SeO_3)_2(SO_4)(H_2O)_2$. The vivid observation of the $^5D_0 \rightarrow ^7F_0$ transition, along with the nuclear spin-free chalcogen constituents, positions $Eu_2(SeO_3)_2(SO_4)(H_2O)_2$ as a potential quantum memory candidate. The full width at half maximum (FWHM) of the $^5D_0 \rightarrow ^7F_0$ transition is 7.42 meV **(Figure S10a)** with an optical lifetime of 3.39 μs and 3.58 μs at 78 K and 298 K, respectively. **(Figure 3d)**. The decay curve was fitted using a triexponential function (equation 9).[45]

$$\frac{I}{I_0} = \sum_{i=1}^{n=3} A_i e^{(t/\tau_i)} \quad (9)$$

Where $I$ and $I_0$ are the photoluminescence intensity as a function of time, $t$, and the incident radiation, respectively, while $A_i$ and $\tau_i$ are the amplitude and lifetime of each decay component, respectively. The multiexponential fit can be attributed to a combination of radiative decays from the $^5D_0$ → $^7F_0$ transition and the nearby $^5D_0 \rightarrow ^7F_1$ transition observed at 588 nm, and the fast phonon-assisted non-radiative decay processes. With the ±10 nm bandpass of the NanoLED pulsed light source, the chances of capturing the nearby transitions in the lifetime measurement becomes substantial. This is also supported by the fitted result (**Table 1**), where the decay having the longest lifetime and the smallest amplitude is attributed to the $^5D_0 \rightarrow ^7F_0$ transition, while the other decay components, potentially derived from the nearby $^5D_0 \rightarrow ^7F_1$ transition and the fast non-radiative decay processes, are associated with the shorter lifetimes. The integrated count rate of the decay curve shows significant consistency at both 78 K and 298 K **(Figure S10)**. The temperature-dependent measurement, performed at 78 K ≤ T ≤ 400 K, reveals no variation in FWHM. We plan to conduct further studies on the optical linewidth and lifetime of the $^5D_0 \rightarrow ^7F_0$ transition at $T \leq 4$ K. **Table 2** summarizes relevant stoichiometric $Eu^{3+}$ compounds and the reported optical lifetime. Some of the reported compounds exhibit optical lifetimes in the millisecond range.[1, 8, 17, 20, 46-49] The difference can be attributed to phonon-assisted non-radiative decay competing with radiative decay. **Figure S9** further supports the polar chiral local symmetry of the $Eu^{3+}$ ion in $Eu_2(SeO_3)_2(SO_4)(H_2O)_2$, as evidenced by the UV (393 nm), blue light (463 nm), and green light (533 nm) excitations, all of which exhibit strong emission intensities.

For solid-state phosphor applications, compounds of $Eu^{3+}$ exhibit robust luminescence properties, emitting characteristic orange to red luminescence when excited with UV radiation.[50, 51] This phenomenon is underlined by the nature of the transitions, namely magnetic dipole transitions, electric dipole transitions or electric quadrupole transitions.[52] Since the dominant emitting state is the $^5D_0$, the emission spectra of $Eu^{3+}$ compounds are commonly characterized by the $^5D_0 \rightarrow ^7F_J$ ($J = 1 – 6$) transitions driven by magnetic dipole interactions and electric dipole interactions.[53-55] The bands associated with the $^5D_0 \rightarrow ^7F_J$ ($J = 5 – 6$) transitions are not commonly observed as they lie beyond the limit of most detectors.[52] The electric dipole transitions originate from the interaction between $Eu^{3+}$ and the electric field component of light, creating a linear flow of charge.[52] Such transitions have odd transformation parity with respect to inversion symmetry, and hence are Laporte forbidden.[44] However, when placed in a crystal lattice, the rules are relaxed by vibronic coupling and orbital mixing arising from the ligand-field effect, spin-orbit coupling, and point symmetry selection rules.[56] The $^5D_0 \rightarrow ^7F_J$ ($J = 2, 4$) transitions are forced electric dipole driven, and the intensity of the $^5D_0 \rightarrow {}^7F_2$ transition dominates in non-centrosymmetric lattices.[57] The magnetic dipole transition $^5D_0 \rightarrow {}^7F_1$ is insensitive to crystal field effects and site symmetry as the intensity of magnetic dipole transition is invariant between free ion and crystal electric field environment. Hence, it is used as a standard for calibrating the intensities of the other emission peaks.[52] The emission spectrum becomes more complex when transitions from higher excited $^5D_J$ states ($^5D_1$, $^5D_2$, $^5D_3$) are observed.[44] However, concentration quenching minimizes these transitions in favor of the lower energy level emission when non-radiative processes such as cross-relaxation dominate.[58] The vibrational energy of the host lattice also discriminates against the higher energy level emissions ($^5D_J$, $J$ = 1, 2, 3) since a high cut-off vibrational energy implies rapid non-radiative relaxation to the lower energy levels.[44,52] On this premise, highly concentrated $Eu^{3+}$ compounds with anions or ligand groups with high vibrational energy, such as $H_2O$, may show photoluminescence emissions primarily from the $^5D_0$ state.[59] Without considering charge transfer bands, $Eu^{3+}$ emission spectrum is often monitored by parking the wavelength at ~393 nm, which is usually the most intense excitation band ($^5L_6 \leftarrow {}^7F_0$).[44]

**Table 1: Fitting results of $Eu_2(SeO_3)_2(SO_4)(H_2O)_2$ decay curve using a triexponential function**

| Parameter | $A_1$ | $A_2$ | $A_3$ | $\tau_1$ (μs) | $\tau_2$ (μs) | $\tau_3$ (μs) |
|---|---|---|---|---|---|---|
| 78 K | 0.15(1) | 0.06(1) | 0.84(1) | 0.33(1) | 3.39(1) | 0.007(1) |
| 298 K | 0.30(1) | 0.11(1) | 0.65(1) | 0.36(1) | 3.58(1) | 0.005(1) |

To determine the excitation energy for our photoluminescence studies, the emission spectrum was recorded by fixing the excitation energy at 3.139 eV ($\lambda$ = 395 nm), obtained from the absorption spectrum. The emission was recorded from $E$ = 2.95 eV ($\lambda$ = 420 nm) to $E$ = 1.72 eV ($\lambda$ = 720 nm). Then using the emission energy at $E$ = 2.013 eV ($\lambda$ = 616 nm) the excitation spectrum was recorded from $E$ = 4.133 eV ($\lambda$ =300 nm) to $E$ = 2.138 eV ($\lambda$ = 580 nm) to obtain the most intense excitation transition, at $E$ = 3.16 eV ($\lambda$ = 393 nm) (**Figure 3**). In the excitation spectra, the direct absorption of photons by the $Eu^{3+}$ ions gave rise to transitions from the ground state, $^7F_0$, to the various excited state levels, observed at 78 K. From the comparison of the spectra at 78 K and 298 K, the hot transitions representing the transitions from the thermally populated $^7F_1$ level to the $^5D_1$ and $^5L_7$ states can be observed in the room temperature spectra. In the emission spectra, only bands associated with the $^5D_0 \rightarrow {}^7F_J$ ($J$ = 0 - 4) transitions were observed, consistent with the expectation for a lattice with high cut-off vibrational energy. The Stark levels splitting agrees with the low site symmetry of $Eu^{3+}$ in $Eu_2(SeO_3)_2(SO_4)(H_2O)_2$. As discussed earlier, the $^5D_0 \rightarrow {}^7F_0$ transition is observed at 2.145 eV (578 nm), with an intensity strong enough to affirm the polar ($C_1$) site symmetry of the $Eu^{3+}$ ion in the $EuO_7(H_2O)$ polyhedra. The $^5D_0 \rightarrow {}^7F_1$ transition is observed at 2.11 eV (588 nm) split into 3 nondegenerate folds, supporting the $C_1$ point symmetry of the $EuO_7(H_2O)$ polyhedra. The hypersensitive $^5D_0 \rightarrow {}^7F_2$ transition governs the red emission reported for $Eu^{3+}$ compounds.[44,52] However, in centrosymmetric lattices, the intensity of $^5D_0 \rightarrow {}^7F_1$ transition dominates over the $^5D_0 \rightarrow {}^7F_2$ transition, giving an asymmetric ratio < 1, and an orange emission.[60] In $Eu_2(SeO_3)_2(SO_4)(H_2O)_2$, the intensity of the $^5D_0 \rightarrow {}^7F_2$ transition at 2.033 – 2.013 eV (610 - 616 nm) is 3 times stronger than the $^5D_0 \rightarrow {}^7F_1$, giving an asymmetric ratio of 2.97. This is evident in the very bright red luminescence of $Eu_2(SeO_3)_2(SO_4)(H_2O)_2$ (**Figure 3c**). In a polar spatial and point symmetry, the $^5D_0 \rightarrow {}^7F_2$ transition is expected to split into five (5) Stark levels. In $Eu_2(SeO_3)_2(SO_4)(H_2O)_2$, the emission band of the $^5D_0 \rightarrow {}^7F_2$ transition is split into two Stark levels at 2.013 eV (610 nm) and 2.033 eV (616 nm), with the peak at 616 nm showing a broad shoulder that reflects an overlap of different Stark levels. The $^5D_0 \rightarrow {}^7F_3$ transition is usually weak as it is forbidden by the Judd-Ofelt theory.[44,61-63] This transition is observed in $Eu_2(SeO_3)_2(SO_4)(H_2O)_2$ as a weak intensity band at 1.91 eV (650 nm). The $^5D_0 \rightarrow {}^7F_4$ electric dipole transition is observed at 1.823 - 1.746 eV (680 – 710 nm). $Eu^{3+}$ materials with an intense $^5D_0 \rightarrow {}^7F_4$ peak have been reported to emit more profound red light.[64] In $Eu_2(SeO_3)_2(SO_4)(H_2O)_2$, the intensity of the $^5D_0 \rightarrow {}^7F_4$ transition is about 20% less than the $^5D_0 \rightarrow {}^7F_2$ transition. Such relatively high intensity of the $^5D_0 \rightarrow {}^7F_4$ has been attributed to result from the behavior of the Judd-Ofelt intensity parameter, $\Omega_\lambda$ ($\lambda$ = 2, 4, 6) in a highly polarizable chemical environment.[43,65] The temperature-dependent emission spectra measured from 78 K $\leq T \leq$ 400 K show no change in the peak energy of the $^5D_0 \rightarrow {}^7F_J$ ($J$ = 1 – 4) transitions (**Figure S11**). The temperature-dependent full-width-at-half-maximum (FWHM) for the $^5D_0 \rightarrow {}^7F_J$ ($J$ = 1 – 4) transitions is relatively constant at low temperatures and in the 35 – 43 meV range (~9 – 12 nm). Such narrow FWHM is essential for phosphor-converted white light-emitting diodes. The temperature-dependent excitation spectrum at $T$ = 78 K and 296 K (**Figure 3a and Figure S9**) shows features of a thermally enhanced phonon-assisted population of the $^7F_1$ state. This can be seen in the increased intensity and broadening of the $^7F_1 \rightarrow {}^5D_1$, $^7F_1 \rightarrow {}^5D_3$, $^7F_1 \rightarrow {}^5G_2$, $^5G_3$ and $^7F_1 \rightarrow {}^5D_4$ transitions at 298 K. This observation is also captured in the heat capacity modelling where the temperature of the Debye, first and second Einstein modes is clearly maximized at $T$ < 300 K. These phonon modes could assist the thermal population of the $^7F_1$ state and enhance excitations from the $^7F_1$ state to the excited state levels. Also, the absorption and emission characteristics of a lanthanide phosphor are influenced by the interaction of the activator ions and the host lattice. Choice matrices are expected to have lower lattice phonon energy, such as in halides and oxides, high radiative transition rates, and good chemical stability.[52,66] It has been demonstrated that lower lattice energies promote emissions from the different $^5D_J$ excited states of $Eu^{3+}$ to yield a global white light rather than a luminous red.[52] The maximum vibrational energy of a host lattice relates to the maximum frequency in the vibrational spectrum. In $Eu_2(SeO_3)_2(SO_4)(H_2O)_2$, the maximum vibrational energy is the O-H stretching mode at 3475 cm$^{-1}$. This mode transfers excess energy to the H-O-H bending mode upon excitation.[35] From our results, we propose that the ultrafast energy relaxation of the O-H vibrational modes and the strong hydrogen bonding in $Eu_2(SeO_3)_2(SO_4)(H_2O)_2$ which couples the $H_2O$ molecule to the $(SeO_3)^{2-}$ group, could facilitate very rapid non-radiative relaxations to the $^5D_0$ state, thereby enhancing the red emissions ($^5D_0 \rightarrow {}^7F_J, J$ = 2, 4). From our qualitative assessment, this attribute could explain the bright red luminescence of $Eu_2(SeO_3)_2(SO_4)(H_2O)_2$. Additionally,

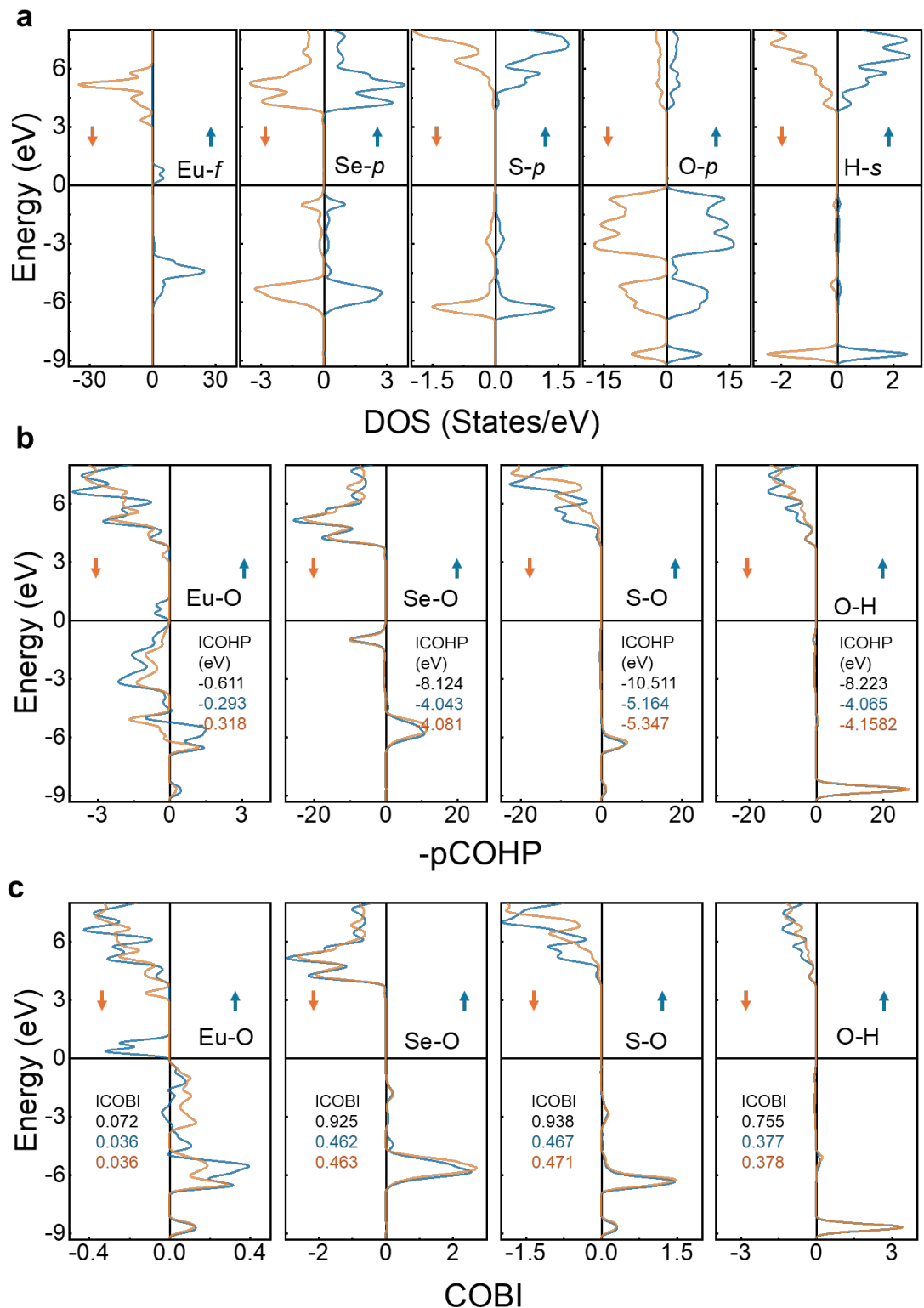

**Figure 4.** **(a)** Density of states plots showing the contributions of the Eu-f, Se-p, S-p, O-p, and H-s interactions in the bonding and antibonding states. The arrows indicate spin-up and spin-down. **(b-c)** Chemical bonding analysis in the framework of crystal orbital Hamilton population (COHP) and crystal orbital bonding index (COBI).

the high concentration of the $Eu^{3+}$ ions with a Eu-Eu distance, 4.0135(4) Å would favor cross-relaxation between neighboring $Eu^{3+}$ ions.[58] This concentration quenching of the high-energy excited states further enhances the ($^5D_0 \to$ $^7F_J$, $J$ = 2, 4) transitions, including weak emissions such as $^5D_0 \to$ $^7F_0$. To justify this proposition and provide evidence for a single-phonon or multiphonon-assisted energy transfer, time-resolved emission spectra would be essential. No charge transfer bands are observed within the limits of our detector (300 – 800 nm). This suggests the charge transfer

**Table 2: Some stoichiometric $Eu^{3+}$ compounds, space group, ligand-field site symmetry of the $Eu^{3+}$, and the wavelength, linewidth and coherence time of the $^5D_0 \rightarrow {}^7F_0$ transition.**

| Material | Space group | Wyckoff symmetry of $Eu^{3+}$ | Ligand-field site symmetry of $Eu^{3+}$ | Eu-Eu distance (Å) | Wavelength (nm) | Optical lifetime, $T_1$ (ms) | Experimental temperature (K) | Reference |
|---|---|---|---|---|---|---|---|---|
| $NaEu(IO_3)_4$ | $Cc$ | 1 | $C_2$ | 5.6082(5) | 579.9 | 1.5 | 1.7 | 1 |
| $EuCl_3 \cdot 6H_2O$ (deuterated and isotopically purified for the $Cl^-$ ion) | $P2/n$ | 1 | $C_2$ | 6.3576(12) | 579.7 | 2.6 | 4 | 17, 68 |
| $[Eu(BA)_4(pip)]$ | $P2_1/n$ | 1 | $C_{2v}$ | 9.6492(9) | 580.37 | 0.54 | 4 | 20 |
| $EuP_5O_{14}$ | $P2_1/c$ | 1 | $C_2$ | 5.1738(6) | 578.5 | 5 | 2 | 46, 69 |
| $[Eu(dpphen)(NO_3)_3]$ | $P2_1/c$ | 1 | $C_{2v}$ | 8.1168(6) | 580.91 | 1.42 | 4.2 | 47 |
| $[Eu(trensal)]$ | $P\bar{3}c1$ | 3.. | $C_{3v}$ | 7.7194(3) | 579.45 | 0.354 | 4.2 | 48 |
| $Eu(HCOO)_3$ | $R3m$ | $3m$ | $C_{3v}$ | 6.2056(2) | 580.20 | 1.57 | 1.4 | 8 |
| $Eu(HCOO)_3 \cdot (HCONH_2)_2$ | $C2$ | 2 | $C_2$ | 6.6405(10) | 580.11 | 1.43 | 1.4 | 8 |
| $Eu(IO_3)_3$ | $P2_1/n$ | 1 | $C_1$ | 4.2119(6) | 580 | 1.6 | 4.2 | 70 |
| $Cs_2NaEuF_6$ | $Fm\bar{3}m$ | $m\bar{3}m$ | $O_{h^*}$ | 6.4651(1) | 578 | - | - | 19 |
| $NaEu(PO_3)_4$ | $P2_1/n$ | 1 | $C_{2v}$ | 5.7175(6) | 579 | 0.0048 | 77 | 49, 71 |
| $Eu_2(SeO_3)_2(SO_4)(H_2O)_2$ | $C2$ | 1 | $C_2$ | 4.0135(4) | 578 | 0.0034 | 78 | This work |

The site symmetry with an asterisk does not theoretically support observation of the $^5D_0 \rightarrow {}^7F_0$ transition. The authors noted the $^5D_0 \rightarrow {}^7F_0$ transition observed in $Cs_2NaEuF_6$ could arise from distortion, oxygen defects or disorder in the $NaEuF_4$ impurity.

luminescence, which refers to the transition of an electron into the 4$f$ energy levels of the $Eu^{3+}$, would be deeper in the UV region, consistent with literature reports on $Eu^{3+}$ where the charge transfer bands are observed in the short wavelength UV (220 – 280 nm).[67] Furthermore, $Eu_2(SeO_3)_2(SO_4)(H_2O)_2$ exhibits excellent thermal stability within the investigated temperature range, 78 K ≤ $T$ ≤ 400 K and experimental conditions. The fluorescence intensity of the $^5D_0 \rightarrow {}^7F_2$ transition is maintained at 90% at room temperature, while the lowest intensity, 78 %, is observed at 350 K. The lifetime of the $^5D_0 \rightarrow {}^7F_2$ transition is observed as 0.67 µs at $T$ = 298 K (Figure S10). For color purity, the Commission Internationale de l'Éclairage (CIE) color coordinates of $Eu_2(SeO_3)_2(SO_4)(H_2O)_2$ (0.65, 0.35) approach the ideal red chromaticity (0.67, 0.33) for the National Television

Standard Committee (NTSC) system (**Figure S13**).[72] The color purity was calculated using the relation:

$$\text{Color purity} = \frac{\sqrt{(x_s-x_i)^2+(y_s-y_i)^2}}{\sqrt{(x_d-x_i)^2+(y_d-y_i)^2}} \quad (10)$$

where ($x_s$, $y_s$), ($x_i$, $y_i$) and ($x_d$, $y_d$) are the coordinates of the sample, illuminant point and dominant wavelength, respectively. According to Equation 1, the color purity is determined to be 95.63%.

**Electronic Structure Calculation.** To gain further insight into the chemical bonding and orbital overlap that give rise to the electronic, magnetic and optical properties of $Eu_2(SeO_3)_2(SO_4)(H_2O)_2$, we performed density of states (DOS) calculations (**Figure 4**) and extracted the COHP and COBI information from the local orbital projection of the augmented plane-wave calculation **(Figure S14)**. The valence band maximum shows large contributions from the O-p orbitals while the conduction band minimum indicates contributions from the Eu-f orbitals. From the photoluminescence excitation spectra, we noted that the charge transfer band, which arises from the $Eu^{3+}-O^{2-}$ bond formation and electron transfer, is in the deep UV region. This suggests that though the density of states analysis shows a probability for an electron transfer from O-p orbital to the Eu-f vacant orbital, the $Eu^{3+}-O^{2-}$ interaction is weak and less stable than in the $(SO_4)^{2-}$, $(SeO_3)^{2-}$, and $H_2O$ groups. Hence, the formation of $Eu^{2+}$, which results in a redshift of the broad charge transfer band, is reduced. This is substantiated by the integrated COHP and COBI values for the Eu-O, Se-O, S-O, and O-H bonding interactions (**Figure S14**). The smaller *i*COHP and *i*COBI values of the Eu-O bonding interaction indicate its lesser stability and more ionic nature than the $(SO_4)^{2-}$, $(SeO_3)^{2-}$, and $H_2O$ functional groups. Additionally, the Eu-f contribution to the conduction band minimum provides further insight into the decreased SHG intensity, as the unfilled f-orbital leads to absorption, thereby confirming the $4f^6$ ground state. Overall, the DFT computation supports the photoluminescence, nonlinear optical, and non-magnetic properties of $Eu_2(SeO_3)_2(SO_4)(H_2O)_2$.

## CONCLUSION

We have demonstrated the design of an air-stable $Eu^{3+}$ stoichiometric phase featuring high emitter density, narrow optical line width and microsecond-scale lifetime of the $^5D_0 \rightarrow {}^7F_0$ transition, attributes desirable for quantum memory applications. Our approach involves deliberate lattice and electronic control to stimulate driving forces towards multiple functionalities. This study shows that the asymmetric ligands provide strong crystal field effects on the $Eu^{3+}$ center, driving strong emissions from the $^5D_0 \rightarrow {}^7F_j$ transitions, including the symmetry forbidden $^5D_0 \rightarrow {}^7F_0$ transition, and enhancing the red luminescence. The $^5D_0 \rightarrow {}^7F_0$ transition shows no change in peak energy and FWHM at our experimental conditions, 78 K ≤ T ≤ 400 K. The FTIR vibrational energies of the ligands provide substantial insights into the emission spectra, while specific heat modelling rationalizes the observations in the photoluminescence excitation spectra. The asymmetric ligands also effectively polarize the crystalline structure, endowing $Eu_2(SeO_3)_2(SO_4)(H_2O)$ with the capability to double the frequency of incident radiation via second harmonic generation. For future work, we plan to characterize the relaxation processes, the linewidth, and the lifetime of the $^5D_0 \rightarrow {}^7F_0$ at T ≤ 4 K. This work will also extend to a study on the influence of concentration on the photoluminescence properties by doping $Eu^{3+}$ into $Ln_2(SeO_3)_2(SO_4)(H_2O)_2$ (Ln = La, Y). Overall, our findings demonstrate that the strategic chemical design of bulk materials can be utilized to create customized frameworks that are tunable for quantum memory implementation.

## AUTHOR INFORMATION

### Corresponding Author

* Thao T. Tran, thao@clemson.edu
Department of Chemistry, Clemson University, Clemson, SC 29634, USA

### Author Contributions

The manuscript was written through contributions of all authors. All authors have given approval to the final version of the manuscript.

### Notes

The authors declare no competing financial interest.

## ACKNOWLEDGMENT

This work was funded by Clemson University. The white-light laser was supported by the Clemson Major Research Instrumentation (CU-MRI) initiative. This work made use of the TCSPC controller and Horiba NanoLED units supported by the Arnold and Mabel Beckman Foundation through a 2023 BYI award and the Camille and Henry Dreyfus Foundation grant 2025 Camille Dreyfus Teacher-Scholar award TC-25-071. X.H. and K.R.K. acknowledge the Arnold and Mabel Beckman Foundation through a 2023 BYI award. We greatly appreciate Dr Colin McMillen for assistance in X-ray diffraction and the Halasyamani group for the SHG measurement.

## ABBREVIATIONS

PXRD, powder X-ray diffraction; ATR-FTIR, attenuated total reflection Fourier transform infrared; PL, photoluminescence; PPMS, physical properties measurement system; TGA, thermogravimetric analysis; DSC, differential scanning calorimetry; BA, benzoylacetonate ; pip, piperidin-1-ium ; dpphen, 2,9-bis(pyrazol-1-yl)-1,10-phenanthroline; trensal, 2,2′,2″-tris(salicylideneimino)triethylamine.